\documentstyle[twocolumn,aps,epsf,prb]{revtex}

\begin{document} 
\draft 

\twocolumn[
\hsize\textwidth\columnwidth\hsize\csname@twocolumnfalse\endcsname

\title{Quantum computing and quantum measurement with mesoscopic 
Josephson junctions} 

\author{D.V. Averin}

\address{Department of Physics and Astronomy, SUNY at Stony Brook, 
Stony Brook, NY 11794-3800} 

\date{\today} 
 
\maketitle 
 
\begin{abstract} 

Recent experimental demonstrations of quantum coherence of the 
charge and flux states of Josephson junctions show that the 
quantum Josephson dynamics can be used to develop scalable quantum 
logic circuits. In this work, I review the basic concepts of 
Josephson tunneling and Josephson-junction qubits, and discuss 
two problems of this tunneling motivated by quantum computing 
applications. One is the theory of photon-assisted resonant flux 
tunneling in SQUID systems used to demonstrate quantum coherence of 
flux. Another is the problem of quantum measurement of charge 
with the SET electrometer. It is shown that the SET electrometer at 
the Coulomb blockade threshold is the quantum-limited detector with 
energy sensitivity reaching $\hbar/\sqrt{3}$ in the resonant-tunneling 
regime. 

\end{abstract} 

\vspace*{5ex}

] 

\section{Introduction}   

Remarkable ability of Bose condensates to create a single quantum 
state occupied with a macroscopically large number of particles 
leads to an unusual situation when a macroscopic object behaves as 
a pure quantum state of a simple quantum mechanical system. This 
behavior is in stark contrast to the usual situation when 
macroscopic objects behave purely classically and do not show any 
collective quantum effects despite the underlying quantum dynamics 
of their microscopic elements. The macroscopic quantum states of 
Bose condensates can exhibit non-trivial quantum dynamics associated 
with the fact that the number of particles in the condensate and 
the phase of its wavefunction are non-commuting quantum variables 
\cite{dir,and}. 

Macroscopic quantum phenomena arising from this dynamics are best 
studied in Josephson tunnel junctions, where the role of bosons is 
played by Cooper pairs of electrons, and both the number of pairs and 
the phase of the wavefunction give well-defined and controlled 
contributions to the junction energy. The ``number'' part of the energy 
comes from the Coulomb charging of the junction capacitance $C$ by 
Cooper pairs and has characteristic magnitude $E_C\equiv (2e)^2/2C$, 
while the pair tunneling across the junction creates the ``phase'' 
term in the energy of magnitude $E_J$. The nature of the quantum 
dynamics of 
a Josephson junction depends on the ratio of these two characteristic 
energies. In large junctions with $E_J\gg E_C$ the phase behaves almost 
classically with quantum effects being ``corrections'' to semiclassical 
dynamics. One of the first quantum effects studied in this regime is 
the macroscopic quantum tunneling of the phase -- see reviews in the 
collection \cite{mqt}. In the opposite case of small junctions with 
$E_J\ll E_C$, the junction dynamics can be interpreted as semiclassical 
dynamics of charge on the capacitance $C$. Characteristic feature 
of the charge dynamics is correlated tunneling of individual 
Cooper pairs one by one between superconducting electrodes of the 
system \cite{set,scz,sct}. 

Macroscopic character of the quantum effects in Josephson junction 
dynamics makes them attractive as the basis for development of 
scalable quantum logic circuits for quantum computation. The regime 
of coherent quantum dynamics in Josephson tunneling can be reached 
in relatively large, micrometer-size junctions that can be 
designed and fabricated with large degree of precision. The 
relevant junction parameters, including 
the energies $E_J$ and $E_C$, reflect the collective properties 
of the superconductors forming the junction and are not 
sensitive to their precise microscopic structure on  
the atomic scale. This makes the Josephson junctions much less 
susceptible to the microscopic disorder unavoidable in the 
present-day fabrication technology, than, e.g., semiconductor 
quantum dots in the regime of coherent electron transport, also 
considered for quantum computing applications \cite{qd}.

The advantage of scalability of Josepshon junction and, in general,  
solid-state qubits and logic circuits comes at a price of 
stronger dephasing and energy relaxation in these systems than in 
other non-solid-state qubits. Nevertheless, recent experimental 
demonstrations of quantum coherent Josephson dynamics both in charge 
\cite{t5} and phase \cite{t6,t7} regime show that the problem of 
dephasing and energy relaxation in Josephson junctions is in 
principle solvable, and stimulate further studies of superconducting 
qubits. 
The aim of this work is to review the basic concepts of quantum 
dynamics of mesoscopic Josephson junctions and qubits, and to 
discuss two specific questions motivated by quantum 
computing applications of Josephson junctions: the theory of 
coherent effects in the photon-assisted resonant flux tunneling, 
and the problem of quantum measurement with SET electrometer.

\vspace{1ex} 

\section{Josephson tunneling in mesoscopic Josephson junctions} 

\vspace{1ex} 

An isolated mesoscopic Josephson tunnel junction is a system of 
two small superconducting islands separated by an insulator layer 
that is sufficiently thin to allow electrons to tunnel between the 
islands. The junction is ``mesoscopic'' if its capacitance $C$ is 
small enough for the charging energy $E_C=(2e)^2/2C$ of a single 
Cooper pair to be sufficiently large. Quantitative description of 
this system is most transparent when the superconducting energy gap  
of the islands is much larger than other characteristic junction 
energies, including charging energy $E_C$ and temperature $T$. In 
this case, quasiparticles can not be excited in the superconductors 
and the junction properties can be understood simply from the 
standard notion that the superconductors are Bose-Einstein 
condensates of Cooper pairs, with Cooper pairs occupying a single 
quantum state in each island. If $n_1$ and $n_2$ are the numbers 
of Cooper pairs in the two superconducting islands, the imbalance 
$n$ of the number of Cooper pairs between the two superconductors, 
$n= (n_1-n_2)/2$, gives the charge $2en$ on the junction capacitance 
$C$. The junction Hamiltonian depends on $n$ through the charging 
energy $(2en)^2/2C-2enV_e$, where $V_e$ is the potential difference 
induced between the two superconducting islands by external 
electric field.
  
Another part of the Hamiltonian comes from the tunneling of Cooper 
pairs between the islands. In accordance with the standard practice, 
we denote the amplitude of this tunneling as $-E_J/2$ and take into 
account that the transfer of a Cooper pair between the islands changes 
$n$ by $\pm 1$. Then, combining the electrostatic and the tunneling 
terms, we get the total Hamiltonian $H$ of the junction in the 
$n$-representation: 
\begin{equation} 
H= E_C (n-q)^2 -\frac{E_J}{2}(|n\rangle \langle n+1| +
|n+1\rangle \langle n|) \, .
\label{n1} \end{equation} 
Here $q\equiv CV_e/2e$ is the charge (in units of $2e$) induced on 
the junction capacitance by the source of potential $V_e$. 
The energy $E_J$ of Cooper pair tunneling is usually referred to as 
Josephson coupling energy. It can be expressed in terms of the 
normal-state tunnel resistance of the junction. 

Behavior of a single Josephson junction described by the 
Hamiltonian (\ref{n1}) is non-trivial due to important 
difference between the two charges $n$ and $q$ that 
define its charging energy. The charge $n$ originates from the total 
number of Cooper pairs in a superconducting island and can take on 
only integer values, as a charge of any isolated conductor. By 
contrast, the charge $q$ is the polarization charge induced on the 
capacitance $C$ by external electric field and can be varied 
continuously. Because of this difference, eigenstates of the 
Hamiltonian (\ref{n1}) form energy bands \cite{t1} periodic in 
$q$ with the period one. When $q$ changes by one, a Cooper pair 
is transferred through the junction increasing $n$ also by one, 
and fully compensating for the increase of $q$, whereas smaller 
changes of $q$ can not be compensated by any transfer of integer 
number of Cooper pairs. 

\begin{figure}[htb]
\setlength{\unitlength}{1.0in}
\begin{picture}(3.,2.1) 
\put(.2,.0){\epsfxsize=2.6in\epsfysize=2.1in\epsfbox{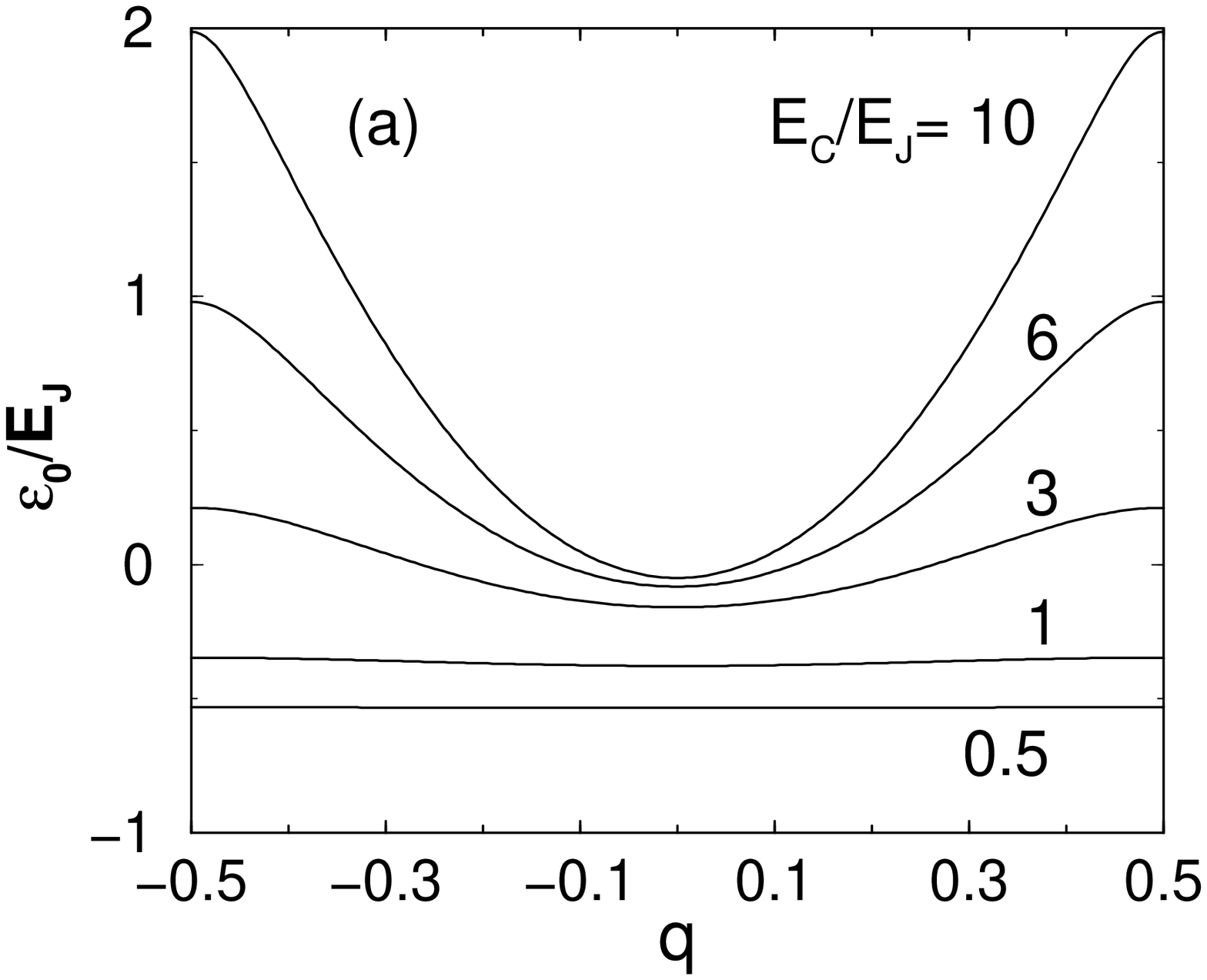}}
\end{picture}
\begin{picture}(3.,2.1) 
\put(.3,.0){\epsfxsize=2.6in\epsfysize=2.1in\epsfbox{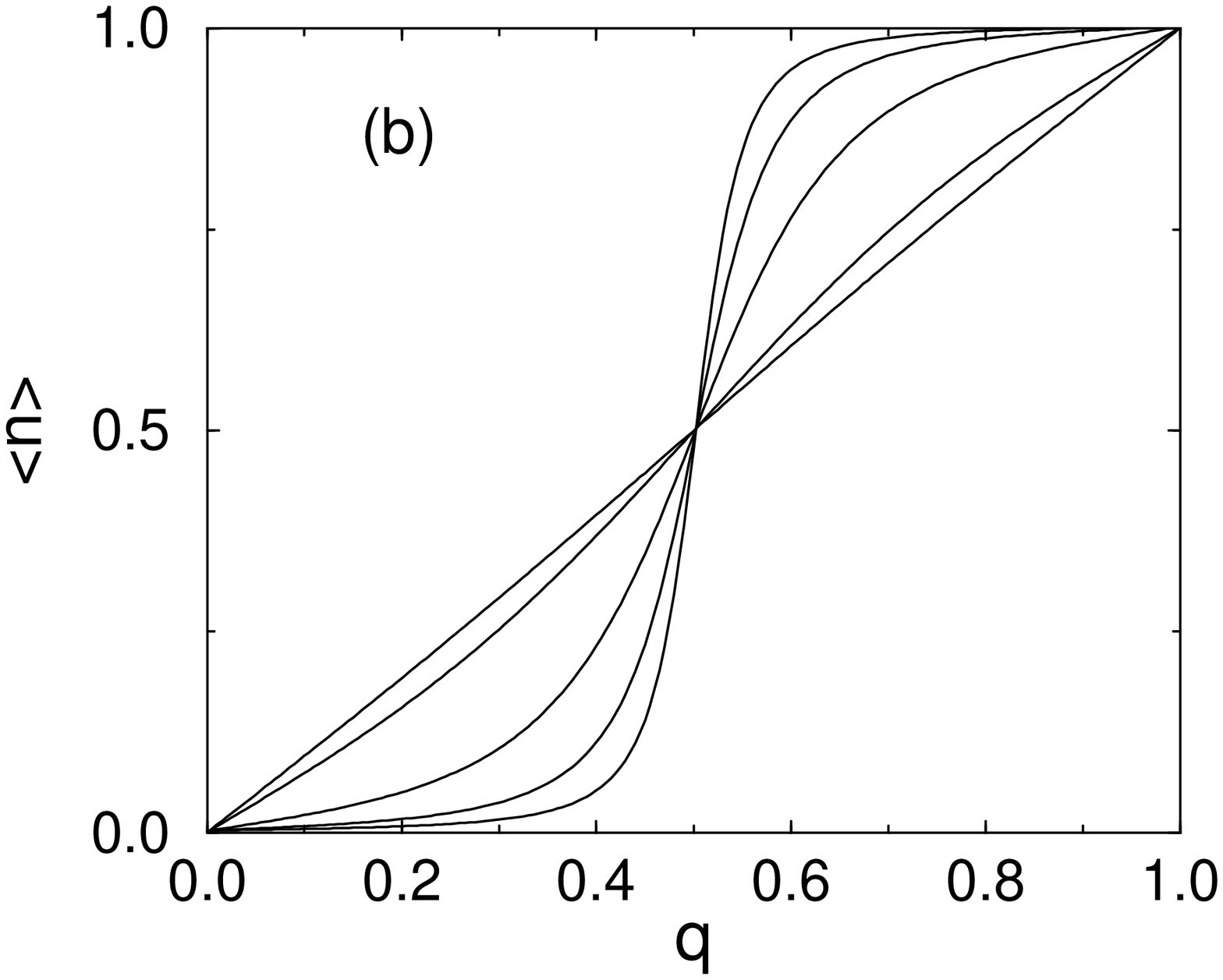}}
\end{picture}
\caption{The ground state energy (a) and the average number of 
Cooper pairs on the junction capacitance (b) for an isolated 
mesoscopic Josephson junction, as functions of the external charge 
$q$ injected into the junction. The curves in (b) are for the 
same values of the ratio $E_C/E_J$ as in (a). }
\label{f1} \end{figure}

An example of the periodic dependence of junction properties on 
$q$ is given in Fig.\ \ref{f1}a which shows one period of 
$q$-dependence of the ground state energy $\varepsilon_0$ of the 
junction 
obtained by direct numerical diagonalization of the Hamiltonian 
(\ref{n1}) for several values of the $E_J/E_C$ ratio. Periodic 
oscillations of the ground-state energy, and similar oscillations 
of energies $\varepsilon_m$ of the excited states, result in 
oscillations of all other junction characteristics with $q$. 
For instance, differentiation of the junction free energy $F= -T 
\ln \sum_m e^{-\varepsilon_m/T}$, with repect to $q$ gives the 
average number of Cooper pairs $\langle n \rangle$ charging the 
junction: 
\begin{equation} 
\langle n \rangle= q - \frac{1}{2E_C} \frac{\partial F}{\partial 
q} \, , 
\label{n2} \end{equation} 
and the average voltage across the junction $V(q)=2e(\langle n 
\rangle -q)/C$. 

Average number of Cooper pairs (\ref{n2}) on the junction 
capacitance as a function of $q$ is shown in Fig.\ \ref{f1}b for 
zero temperature and different $E_J/E_C$ ratios. In the limit of 
small coupling energies $E_J$, the number of pairs $\langle n 
\rangle$ increases in a step-wise manner remaining constant for 
almost all of $q$ values exept for half-integer $q$'s, when it rapidly 
increases by one. Qualitatively, at $E_C\gg E_J$, the supercurrent 
flow between the junction electrodes is ``discretized'' into the 
transfer of individual Cooper pairs by strong Coulomb repulsion, 
and variations of the charge $q$ injected into the junction leads 
to the possibility of controlling their tunneling.  

Step-like behavior of $\langle n \rangle$ at $E_J\ll E_C$ 
reflects quantization of the 
number of Cooper pairs in the superconducting electrodes of the 
junction. When $E_J$ increases, the junction electrodes become 
strongly coupled and the quantization of $n$ is gradually washed 
out. In this case, the ground state energy of the junction (Fig.\ 
\ref{f1}a) becomes virtually independent of $q$, and the average 
number $\langle n \rangle$ of Cooper pairs on the junction 
follows closely the induced charge $q$ (Fig.\ \ref{f1}b),  
$\langle n \rangle \simeq q$. Qualitatively, 
this suppression of $n$ quantization with increasing $E_J$ is the 
result of large quantum fluctuations of $n$ induced by tunneling 
terms in the Hamiltonian (\ref{n1}). 

At $E_J\geq E_C$, the junction properties can be described more 
directly by introducing the phase operator $\varphi$ canonically 
conjugated to the number $n$ of Cooper pairs. From quantum 
mechanics of a single boson mode \cite{dir}, it is known that 
the operator of the phase $\varphi_j$ of the condensate 
wavefunction in one of the junction electrodes, $j=1,2$, can be 
introduced through the operators $s^{\pm} =|n\pm 1\rangle \langle n|$ 
(omitting the index $j$ to simplify the notations) raising and 
lowering the number of Cooper pairs in the condensate. On all of 
the states $|n\rangle$ besides the state $|0\rangle$ with no 
particles, the operators $s^{\pm}$ satisfy the relations 
$s^+s^-=s^-s^+=1$ and $s^-=(s^+ )^\dagger$ which suggest that they 
can be expressed as 
\begin{equation}
s^{\pm}=e^{\pm i\varphi}\, . 
\label{n3} \end{equation}
In fact, since the state $|0\rangle$ is annihilated by $s^-$, 
$s^- |0\rangle =0$, the commutator of $s^+$ and $s^-$ does not 
vanish:    
\begin{equation} 
[ s^-,s^+]= |0\rangle \langle 0| \, .   
\label{n4} \end{equation} 
Non-vanishing commutator contradicts the possibility of expressing 
the operators $s^{\pm}$ through one phase operator, 
and leads to the well-known problem with a generic definition of 
the phase of a boson mode, discussed in the literature on quantum 
optics  -- see, e.g., the review \cite{ph} and references therein. 

However, in the case of a Bose condensate with macroscopically 
large number of particles, all relevant states do not have the 
component $|0\rangle$, the commutator (\ref{n4}) effectively vanishes, 
and eqs.\ (\ref{n3}) define the operators $e^{\pm i\varphi}$. 
Then, the commutator between $n$ and the raising/lowering operators 
can be written as 
\begin{equation}
e^{\pm i \varphi} (n \pm 1) = n e^{\pm i \varphi}  \, .
\label{n5} \end{equation} 
For a superconducting island, the number of Cooper pairs in the 
condensate can be estimated as the ratio $\Delta/\delta$ of the 
superconducting energy gap $\Delta$ over the spacing $\delta$ of the 
single-particle energy levels. Condition $\Delta \gg \delta$ necessary 
for defining the phase operator coincides with the criterion of the BCS 
superconductivity \cite{and2,ml}. 

With the raising and lowering operators $e^{\pm i \varphi_j}$ 
for the two condensates in the junction electrodes, the operator of 
transfer of a Cooper pair through the junction can be written as 
$e^{\pm i \varphi}\equiv e^{\pm i (\varphi_1-\varphi_2)}$. They have 
the same commutators (\ref{n5}) with the number of Cooper pairs $n$ 
on the junction capacitance $C$. The operators $e^{\pm i \varphi}$ 
allow one to define any periodic function of $\varphi$, and the 
commutation relations (\ref{n5}) imply that in this periodic 
$\varphi$-representation, the number of Cooper piars can be written 
as $n= -i \partial/\partial \varphi$. (The aperiodic operator of the 
phase $\varphi$ itself can not be defined for an isolated junction, 
since $n$ is integer and all junction properties should be 
periodic in $\varphi$ with the period $2\pi$). The Hamiltonian 
(\ref{n1}) is equivalent then to that of a plain rotor in an 
external field: 
\begin{equation} 
H= E_C (\frac{1}{i}\frac{\partial}{\partial \varphi}-q)^2 -E_J 
\cos \varphi \, .
\label{n6} \end{equation} 
This representation of the Hamiltonian (\ref{n1}) allows one to 
analyze simply the behavior of the ground state energy $\epsilon_0$ 
and average charge $\langle n\rangle$ (Fig.\ \ref{f1}) for 
$E_J\geq E_C$ in terms of tunneling through the coupling potential 
$-E_J \cos \varphi$.  

Descrpition in terms of the phase dynamics becomes particular relevant 
in the situation of an ``rf SQUID'' (see Fig.\ \ref{f6} below),  
when the junction is included in external superconducting loop. 
Due to quantization of the magnetic flux $\Phi$ in the loop (see, 
e.g., \cite{flux}) the phase $\varphi$ is linked to the flux, 
$\varphi=2\pi\Phi/\Phi_0$, where $\Phi_0=\pi \hbar/e$ is the 
``flux quantum''. The states of this system that differ in 
phase by $2\pi$ are clearly distinguishable, and the charge 
$n=-i\partial/\partial \varphi$ is no loger integer. The 
transition from periodic to aperiodic phase $\varphi$ in the 
Josephson junctions has been debated extensively in the early 
discussions of charging effects in Josephson junctions 
\cite{t1,p1,p2,p3}. In physical terms, in the rf SQUID, both 
electrodes of the junction are parts of the same conductor. In 
this situation, the charge $n$ is the polarization charge and 
can vary continuously. One consequence of this is that any 
external charge $q$ can be screened by charge transfer through 
the loop, so that $q$ does not have any effect on the 
junction dynamics and can be omitted from the Hamiltonian. 
Then, the Hamiltonian of the system, including the 
magnetic energy of the loop inductance $L$ threaded by an 
external flux $\Phi_e$ is: 
\begin{equation} 
H= E_C n^2 -E_J \cos \varphi + \frac{(\Phi_0 \varphi/2\pi- 
\Phi_e)^2 }{2L} \, .
\label{n7} \end{equation} 
In such an ``aperiodic'' system, the phase operator $\varphi$ is 
well-defined and has the standard commutation relation with the 
charge $n$:\cite{and}
\begin{equation}
[\varphi,n]=i \, .
\label{n8} \end{equation} 

Both the controlled quantum dynamics of individual Cooper pairs 
governed by the Hamiltonian (\ref{n1}) and quantum dynamics 
of flux (\ref{n7}) can be used to build qubits for quantum 
computation. The main obstacle to this is dephasing and relaxation 
mechanisms present in the Josephson junction circuits that couple 
charge or flux motion to other degrees of freedom not included in 
eqs.\ (\ref{n1}) or (\ref{n7}). Such a coupling makes 
dynamics irreversible and suppresses quantum correlations of the 
charge or flux states. Although suppression of the relaxation 
strength to acceptably low level is a challenging problem, 
reversible quantum coherent dynamics has been demonstrated both
in the charge and flux regimes. In the charge case, this has been 
done implicitly \cite{t2,t3,t4}, through observation of 
splitting of the two lowest energy eigenstates of the Hamiltonian 
(\ref{n1}) at $q\simeq 1/2$, and explicitly, by direct 
measurement of the time-dependent oscillations of a Cooper pair 
\cite{t5}. Quantum coherence of the flux states remained elusive 
for many years. Very recently, two experiments \cite{t6,t7} 
reported first observations of the coherent superposition of the 
macroscopically distinct flux state, with superposition of states 
evidenced indirectly by the coherent splitting of the energy 
eigenstates of the Hamiltonian similar to (\ref{n7}). The 
splitting is caused by tunneling through the maximum of the SQUID 
potential $U(\Phi)$ for the flux dynamics. In the next section, I 
discuss briefly the theory of flux tunneling in one of these 
experiments. More detailed presentation of this theory can be 
found in \cite{t9}.

\section{Quantum coherence of flux states} 

\begin{figure}
\setlength{\unitlength}{1.0in}
\begin{picture}(3.,2.4) 
\put(.0,.0){\epsfxsize=3.in\epsfysize=2.4in\epsfbox{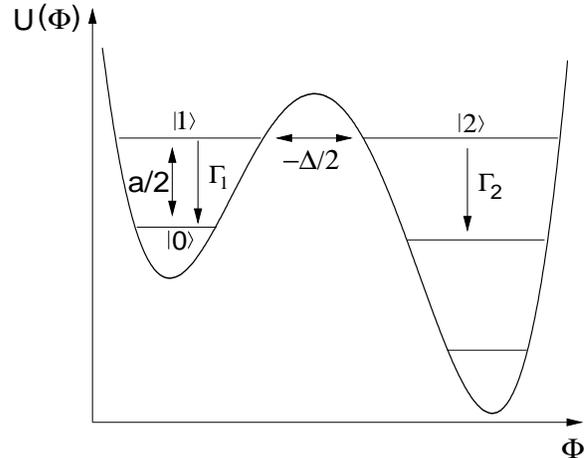}}
\end{picture}
\caption{Schematic diagram of the ``photon-assisted'' 
macroscopic resonant tunneling of flux stimulated by an rf 
perturbation of strength $a/2$. }  
\label{f2} \end{figure}

With the appropriate choice of parameters of the SQUID Hamiltonian 
(\ref{n7}), $\Phi_e \simeq \Phi_0/2$ and $E_J\simeq \Phi_0^2/2\pi^2L$, 
the SQUID potential $U(\Phi)$ has two minima separated by a 
potential barrier. Quantum coherence of flux state should manifest 
itself as coherent oscillations of flux between the two potential 
wells due to tunneling through the barrier. The evidence for such 
coherent oscillations was obtained in \cite{t6} by studying 
photon-assisted resonant tunneling of flux between the two well. 
The energy diagram of this process is shown in Fig.\ 
\ref{f2}.\cite{t8} An rf signal of frequency $\omega$ resonantly 
couples the ground state $|0\rangle$ in the left well of the 
potential to an excited 
state $|1\rangle$ with energy $E$ in this well. The condition of the 
resonant excitation is that the detuning $\nu=E-\omega$ is small, 
$\nu \ll E$. If the amplitude $a$ of the rf excitation is also 
relatively small, $a\ll E$, photon-induced coupling between 
the states $|0\rangle$ and $|1\rangle$ can be described in the 
rotating-wave approximation, in which the terms in the coupling 
Hamiltonian that oscillate rapidly in time are neglected. 
The excited state $|1\rangle$ in the left well is coupled 
resonantly with amplitude $-\Delta/2$ to a state $|2\rangle$ in the 
right well that is shifted in energy by $\varepsilon$ with respect 
to $|1\rangle$. The total Hamiltonian for the flux dynamics in the 
basis of the three states $|0\rangle$, $|1\rangle$, and $|2\rangle$ 
is:  
\begin{equation} 
H=  \left( \begin{array}{ccc} 0 & a/2 & 0 \\ 
a/2 & \nu &  -\Delta/2\\ 0 & -\Delta/2 & \nu -\varepsilon 
 \end{array} \right) \, . 
\label{p2} \end{equation}  

The rate $\tau^{-1}$ of resonant flux tunneling between the wells 
of the SQUID potential in absence of rf signal \cite{t11,t12,t13} 
is known to be qualitatively independent of the strength of the 
energy relaxation in the flux dynamics. The shape of the resonant 
tunneling peaks is the same both in the regimes of 
coherent and incoherent tunneling. In contrast to this, the rate 
of the photon-assisted tunneling is sensitive to the relaxation 
strength, with the resonant peaks acquiring additional structure in 
the regime of coherent tunneling, when the energy relaxation is 
weak. 

The main source of the energy relaxation for quantum dynamics of 
Josephson junctions is electromagnetic environment of the junction. 
Under the assumption that electromagnetic modes of the environment 
are in equilibrium at temperature $T$, the interaction between the 
flux $\Phi$ and the heat bath of these modes can be written as 
\begin{equation}
V=-I_f \Phi \, ,  
\label{p3} \end{equation} 
where $I_f$ is the fluctuating current created in the SQUID 
loop by the environment, with the correlation function given 
by the fluctuation-dissipation theorem: 
\begin{equation} 
\langle I_f(t) I_f(t+\tau) \rangle = \int \frac{d \omega }{\pi} 
\frac{\omega G(\omega) e^{i\omega \tau } }{1-e^{- \omega /T} } \, . 
\label{p4} \end{equation}
Here the brackets $\langle ... \rangle$ denote averaging over 
the equilibrium density matrix of the environment, and $G(\omega)$ 
is the dissipative part of the environment conductance. 

The coupling (\ref{p3}) leads to the two types of relaxation processes 
for dynamics of flux tunneling in the double-well potential. One is 
the ``intrawell'' relaxation inducing transitions from the states 
$|1\rangle$ and $|2\rangle$ to the states with lower energies 
within the same well. In these transitions, the environment is 
absorbing large energy quanta on the order of energy separation 
between the states in the wells. Another is ``interwell'' relaxation 
which arises from the fluctuating energy shifts of one well relative 
to another caused by the coupling to environment (\ref{p3}). 
Interwell transitions transfer to the environment much smaller energy 
on the order of energies in the Hamiltonian (\ref{p2}). 

The dissipative dynamics of flux with the Hamiltonian (\ref{p2}) can 
be described with the time-evolution equations for the density matrix 
$\rho$ of the system. These equations can be obtained in the standard 
way from the environment characteristics (\ref{p3}), (\ref{p4}) in the 
case of weak relaxation -- see, e.g., \cite{t10}. Under the 
natural assumption that the average flux in the states $|0\rangle$ and 
$|1\rangle$ in the left well of the SQUID potential is the same, 
the part of the coupling (\ref{p3}) that corresponds to the interwell 
relaxation is: 
\begin{equation}
V=-I_f \delta \Phi \left( \begin{array}{ccc} 1 & 0 & 0 \\ 
0 & 1 & 0 \\ 0 & 0 & -1 \end{array} \right) \equiv -I_f \delta 
\Phi U\, ,  
\label{p5} \end{equation} 
where $\delta \Phi$ is half of the difference between the average flux 
in the left and right wells.
While eq.\ (\ref{p5}) is written in the flux basis $|0\rangle\, , 
|1\rangle\, ,|2\rangle$, weak relaxation is conveniently described in 
the basis of the eigenstates $|n\rangle$ of the Hamiltonian (\ref{p2}). 
In this basis, the contribution of the weak interwell relaxation 
(\ref{p5}) to the evolution of the density matrix $\rho$ is given by the 
standard expressions: 
\begin{eqnarray}
\dot{\rho}_{nn} & = &  \sum_m ( \gamma_{mn} \rho_{mm} - 
\gamma_{nm} \rho_{nn}) \, ,\label{p6}  \\  
\dot{\rho}_{nm} & = &  -[\gamma'_{mn} + \frac{1}{2} 
\sum_k ( \gamma_{nk} + \gamma_{mk}) ] \rho_{nm} \, , \;\;\; 
n\neq m \, . \nonumber  
\end{eqnarray} 
Transition and dephasing rates in these equations are: 
\[ \gamma_{nm} = \frac{g |U_{nm}|^2 (\varepsilon_n- 
\varepsilon_m)}{1- e^{-(\varepsilon_n-\varepsilon_m) /T}} \, ,  
\;\;\;\; \gamma'_{nm} =  \frac{gT  }{2} 
(U_{nn}-U_{mm})^2\, , \]
where $U_{nm}$ are the matrix elements of the operator $U$ (\ref{p5}) 
in the eigenstates basis, $\varepsilon_n$ are the eigenenergies, and 
dimensionless parameter $g=2G(\delta \Phi)^2/\hbar$ characterizes 
the strength of the interwell relaxation.

The intrawell relaxation has simple form directly in the flux basis. 
Equations for the off-diagonal elements of $\rho$ in this basis are:
\begin{eqnarray} 
\dot{\rho}_{01} & = & (i \nu - \Gamma_1/2 )\rho_{01} +
ia(\rho_{00}-\rho_{11})/2 -i\Delta \rho_{02}/2  \, , 
\nonumber \\   
\dot{\rho}_{12} & = & -(i \varepsilon + (\Gamma_1+\Gamma_2)/2 ) 
\rho_{12} + i\Delta(\rho_{22}-\rho_{11})/2 -ia \rho_{02}/2  \, ,
\nonumber \\ 
\dot{\rho}_{02} & = & (i (\nu- \varepsilon) - \Gamma_2/2 ) 
\rho_{02} - ia\rho_{12}/2 -i\Delta \rho_{01}/2  \, . \label{p7}   
\end{eqnarray} 
The rates $\Gamma_{1,2}$ in these equations describe, respectively, 
the relaxation out of the states $|1\rangle$ and $|2\rangle$ into the 
lower-energy states within the left and right wells. Diagonal part 
of the evolution equations is:  
\begin{eqnarray}
\dot{\rho}_{00} & = &  - a\mbox{Im} \rho_{01} + 
\Gamma_1\rho_{11} + \Gamma_2\rho_{22} \, ,\nonumber \\  
\dot{\rho}_{11} & = &  a\mbox{Im} \rho_{01} + \Delta 
\mbox{Im} \rho_{12} -\Gamma_1\rho_{11} \, , \label{p8} \\  
\dot{\rho}_{22} & = &  - \Delta \mbox{Im} \rho_{12}  - 
\Gamma_2\rho_{22} \, . \nonumber  
\end{eqnarray} 
Equation for $\rho_{00}$ here was written under the assumption that after 
the relaxation in the right well the system is returned instantly to 
the initial state $|0\rangle$. This modification of the evolution 
equations allows us to calculate the rate $\tau^{-1}$ of the flux 
tunneling between the two wells from the stationary solution 
$\rho^{(0)}$ of the evolution equations\cite{t9}: 
\begin{equation}  
\tau^{-1} = \Gamma_2\rho^{(0)}_{22} \, . 
\label{p9}  \end{equation}

The evolution equations for $\rho$ that include both the interwell 
and intrawell relaxation can be solved numerically. To do this, one 
diagonalizes the 
Hamiltonian (\ref{p2}), calculates the interwell relaxation terms 
(\ref{p6}) in the eigenstates basis and transforms them into the flux 
basis where the the intrawell relaxation has the simple form (\ref{p7}), 
(\ref{p8}). Finally, the stationary value of the density matrix $\rho$ 
gives the flux tunneling rate (\ref{p9}). 

\begin{figure}[htb]
\setlength{\unitlength}{1.0in}
\begin{picture}(3.0,2.3) 
\put(0.,0.){\epsfxsize=3.0in\epsfysize=2.3in\epsfbox{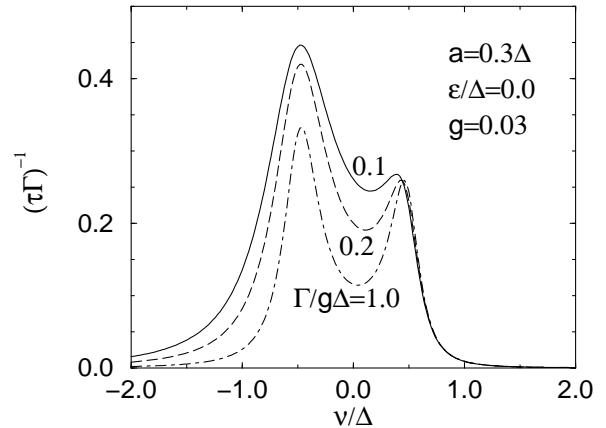}}
\end{picture}
\caption{The rate $\tau^{-1}$ of the photon assisted flux tunneling 
as a function of the detuning $\nu$ in the case of symmetric 
coupling between the tunneling flux states, $\varepsilon=0$, in 
presence of both the interwell and intrawell relaxation. Different 
curves correspond to different magnitude of the intrawell relaxation 
rate $\Gamma$ relative to the interwell relaxation. }
\label{f3} \end{figure}

Figures \ref{f3} shows an example of the results of such a 
calculation obtained for vanishing temperature $T$ and $\Gamma_1 = 
\Gamma_2 \equiv \Gamma$. The tunneling rate is plotted as a function 
of the detuning $\nu$ for $\varepsilon=0$ and several values of the 
relative strength of the intrawell relaxation $\Gamma$. We see that 
the main resonant tunneling peaks is split in the two peaks due to 
coherent oscillations of flux between the two wells. Appearance of 
such 
splitting can be easily understood, since sufficiently weak rf signal 
excites the system not into the state $|1\rangle$ localized in the 
left well, but into the two hybridized states formed out of the states 
$|1\rangle$ and $|2\rangle$ in the two wells. The two different 
energies of the two hybridized states lead to the two peaks in the 
tunneling rate. In the time-domain, hybridization of the states 
$|1\rangle$ and $|2\rangle$ corresponds to the coherent oscillations 
of flux between these states. This means that the splitting of the 
resonant tunneling peak is the direct manifestation of the 
quantum coherent oscillations of flux between the two wells of 
the SQUID potential.

At $\varepsilon=0$, the hybridized states are distributed 
symmetrically between the two wells, and asymmetry of the two peaks 
in Fig.\ \ref{f3} is caused by the interwell relaxation. The 
positive-$\nu$ side of the double-peak 
structure corresponds to excitation of the system into the 
lower-energy eigenstate and is unaffected by the interwell 
relaxation at zero temperature, since there is no energy in this 
regime to create additional tunneling path. In contrast, the 
negative-$\nu$ side of the double-peak structure corresponds to 
excitation of the system into the eigenstate with larger energy, 
and the interwell relaxation increases the rate of tunneling out 
of this state. Because of this, the negative-$\nu$ peak in Fig.\ 
\ref{f3} is larger that the tunneling peak at positive $\nu$, and 
the tunneling rate at $\nu<0$ decreases much more slowly away from 
the peak than at positive $\nu$. 

\begin{figure}[htb] 
\setlength{\unitlength}{1.0in}
\begin{picture}(3.0,2.3) 
\put(0.,0.){\epsfxsize=3.0in\epsfysize=2.3in\epsfbox{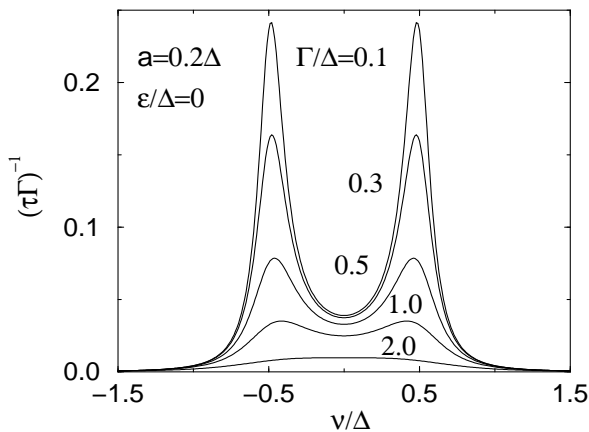}}
\end{picture}
\caption{Evolution of the double-peak structure in the rate 
$\tau^{-1}$ of the photon-assisted resonant flux tunneling as a 
function of detuning $\nu$ with increasing relaxation rate 
$\Gamma$. }  
\label{f4} \end{figure}

In absence of the interwell relaxation, the double-peak structure 
is symmetric for $\varepsilon=0$ as illustrated in Fig.\ \ref{f4}. 
This figure also shows how the splitting of the tunneling peak is 
suppressed with increasing relaxation strength. The curves are 
obtained by finding the stationary solution of eqs.\ (\ref{p7}) and 
(\ref{p8}) which are valid for arbitrary strength of the intrawell 
relaxation, $\Gamma_{1,2} \sim \Delta, a$, and neglecting the 
weak interwell relaxation (\ref{p6}) in this regime. They show 
that for $\Gamma \geq \Delta $ the two peaks of the double-peak 
structure in the flux tunneling rate are suppressed and merge into 
one broad resonant tunneling peak. 

It should be noted that for sufficiently large bias $\varepsilon$, 
the two peaks of the resonant flux tunneling exist even for strong 
relaxation rate $\Gamma$, as can be seen analytically in the limit 
$\Gamma_{1,2} \gg \Delta, a$. Expressing the off-diagonal element of  
$\rho$ in terms of the diagonal ones in the stationary regime from 
eqs.\ (\ref{p7}) and using their stationary values in eqs.\ 
(\ref{p8}) we find the stationary probability $\rho_{22}^{(0)}$ and 
the flux tunneling rate (\ref{p9}). For strong relaxation, 
\begin{equation}
\tau^{-1} = \frac{\Gamma_2 a^2 \Delta^2}{
(4\nu^2+\Gamma_1^2) (4(\nu-\varepsilon)^2 +\Gamma_2^2) } \, . 
\label{p10} \end{equation}
At large bias energy $\varepsilon$, the tunneling rate (\ref{p10}) 
as a function of detuning $\nu$ contains two separate peaks that 
correspond to excitation of the flux into the energy states 
$|1,2\rangle$ localized in the two wells. 
The two peaks merge, however, with decreasing bias and form one broad 
peak at $\varepsilon \simeq 0$ shown in Fig.\ \ref{f4}. This behavior 
should be contrasted with the coherent-tunneling regime with small 
$\Gamma$, when the two peaks remain separated even at $\varepsilon 
\simeq 0$. 

In the experiment\cite{t6}, the flux tunneling rate $\tau^{-1}$ was 
studied as a function of the external flux $\Phi_e$ through the SQUID 
loop. Variations of $\Phi_e$ change both the energy shift 
$\varepsilon$ between the two wells of the SQUID potential $U(\Phi)$ 
and the separation $E$ between the levels within the same well. For 
a fixed frequency $\omega$ of external rf radiation, detuning 
$\nu=E-\omega$ changes together with $E$. Although there are some 
minor quantitative differences between the dependence of the tunneling 
rate $\tau^{-1}$ on $\Phi_e$ (which controls both $\nu$ and 
$\varepsilon$), and dependence of $\tau^{-1}$ on the detuning $\nu$ 
plotted in Figs.\ \ref{f3} and \ref{f4}, both dependences show 
the splitting of the resonant tunneling peaks due to coherent 
superposition of the flux states in the two wells of the potential.    
This means that observation\cite{t6} of the splitting of the 
resonant peaks of the photon-assisted flux tunneling under the resonant 
bias conditions $\varepsilon \simeq 0$ demonstrates quantum coherent 
flux dynamics and quantum superposition of macroscopically distinct 
flux states. This coherent flux dynamics can be used for development 
of the flux qubits. 

\section{Charge and flux qubits} 

\begin{figure}[htb]
\setlength{\unitlength}{1.0in}
\begin{picture}(3.0,2.2) 
\put(0.,0.){\epsfxsize=3.0in\epsfysize=2.2in\epsfbox{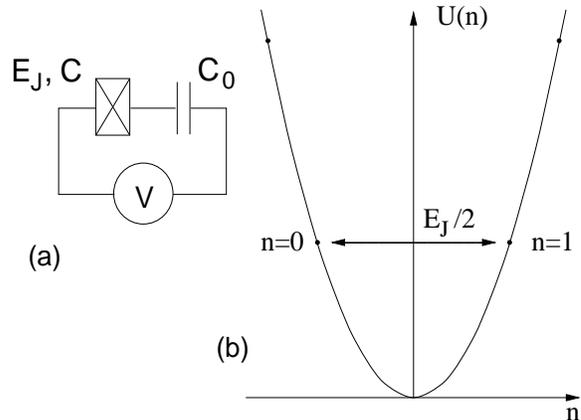}}
\end{picture}
\caption{Equivalent circuit (a) and the energy diagram (b) 
of the charge qubit. External gate voltage $V$ coupled  through 
a small capacitance $C_0$ induces charge close to half a Cooper 
pair charge $2e$ on the Josephson junction with coupling energy 
$E_J$ and capacitance $C$.}  
\label{f5} \end{figure}

Josephson junction qubits and quantum logic circuits can operate in 
the charge and flux regimes. To create a qubit, one needs to reduce 
the junction dynamics to two 
states and to be able to control accurately the resulting two-state 
dynamics. The most basic arrangement that satisfy these requirements 
in the charge space is the single-Cooper-pair box \cite{cb1,cb2,s10} 
configuration, in which the external charge $q$ is injected into 
the junction by a gate voltage through an additional capacitance 
(Fig.\ \ref{f5}a). 

The {\em charge} qubit is the Cooper-pair box with small Josephson 
coupling energy $E_J\ll E_C$ and induced charge $q\simeq 
1/2$.\cite{ssc,sch} In this case, the two charge states, 
$n=0$ and $n=1$, are nearly-degenerate and are separated from 
all other states by the large energy gaps on the order 
of $E_C$ (Fig.\ \ref{f5}b). The two states are coupled by the 
tunneling of one Cooper pair, and the junction Hamiltonian 
(\ref{n1}) is effectively reduced to the two-state Hamiltonian: 
\begin{equation} 
H = E_C (q-\frac{1}{2})(|0\rangle \langle 0| -|1\rangle 
\langle 1|) - \frac{E_J}{2}(|0\rangle \langle 1| +
|1\rangle \langle 0|) \, .
\label{n} \end{equation} 
The energy difference between the charge states is controlled 
through $q$ by the gate voltage, while their tunnel coupling can be 
controlled by the external magnetic field that modulates the 
Josephson coupling energy $E_J$. Operation of a Cooper-pair box as a 
qubit has been demonstrated experimentally in \cite{t5}.

\begin{figure}[htb]
\setlength{\unitlength}{1.0in}
\begin{picture}(3.0,1.95) 
\put(0.,0.1){\epsfxsize=3.0in\epsfysize=1.9in\epsfbox{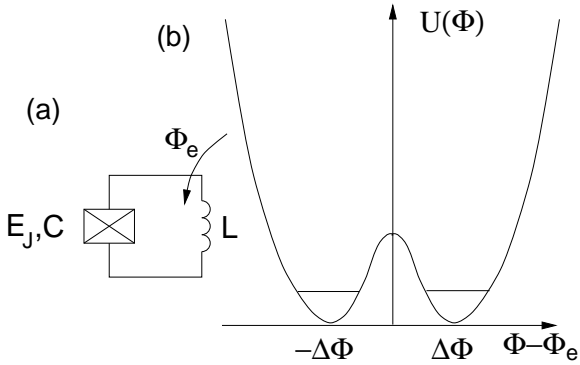}}
\end{picture}
\caption{Equivalent circuit (a) and the energy diagram (b) 
of the flux qubit. External flux $\Phi_e$ induces the flux 
through the inductance $L$ of the rf SQUID that is close to 
half a flux quantum.}  
\label{f6} \end{figure}

The basic scheme of the {\em flux} qubit\cite{fl1,fl2,fl3} 
(Fig.\ \ref{f6}) is similar to the charge qubit. When external 
flux through an rf SQUID (\ref{n7}) is equal to a half of the 
magnetic flux quantum $\Phi_0$, and $E_J\simeq \Phi_0^2/2\pi^2L$, 
the SQUID potential $U(\Phi)$ has two symmetric minima at $\Phi=\Phi_0 
/2 \pm \Delta \Phi$ with the two degenerate ground states in them. 
These two states are separated from the excited states by large energy 
gaps on the order of quantum of oscillation frequency $\omega_p$ 
around the minima, $\hbar \omega_p \simeq (E_JE_C)^{1/2}$, and 
are coupled by tunneling through the barrier separating the minima.
The tunneling amplitude $\Delta/2$ is small, $\Delta/2 \ll 
\hbar \omega_p$. In this 
regime, the SQUID dynamics reduces to the two-states dynamics at 
temperatures $T\ll \hbar \omega_p$. The energy difference 
$\varepsilon$ between the two flux states is controlled by the 
deviations of the external flux $\Phi_e$ from $\Phi_0/2$, 
$\varepsilon \simeq (\Phi_e-\Phi_0/2) \Delta \Phi/L$, while the 
tunneling amplitude can be controlled by additional flux 
modulating the Josephson coupling energy $E_J$ \cite{t12}. This means 
that if the rate of the environment-induced decoherence of the flux 
states is suppressed significantly below the tunneling 
frequency $\Delta$, as in experiments \cite{t6,t7}, the rf SQUID can 
be used as the flux qubit.  

After individual qubits, the next step in the design of the 
quantum computer is the organization of the qubit dynamics 
necessary to perform quantum logic operations. The main general 
requirement to this dynamics is that it should 
maintain the level of precision sufficient for quantum 
computation. For instance, the amplitudes of the unwanted 
transitions to the higher-energy states of the system (leading  
to leakage of the probability out of the two-state qubit 
space\cite{leak} and destroying correct operation of the qubit)   
should be low. The best and probably the only type of dynamics 
that satisfies this requirement in the solid-state systems is 
the adiabatic dynamics \cite{ssc}. When all qubit parameters 
are changed adiabatically, the amplitudes of all unwanted 
transitions decrease exponentially with the ratio of the 
characteristic qubit energies ($E_J$ for the charge and $\Delta$ 
for the flux qubits) to the frequency of qubit operation. For 
comparison, the amplitude of the unwanted transitions in the 
regime of Rabi oscillations employed in the ion-trap \cite{it} 
or NMR qubits -- see, e.g. \cite{nmr}, decreases only linearly 
with the operation frequency/energy ratio. This means that 
despite the term ``adiabatic'', for a given level of accuracy 
and given qubit energies, adiabatic manipulations of the qubit 
states can be performed considerably faster than Rabi 
transitions. Larger operation frequency of adiabatic quantum 
logic gates is particularly important for solid-state gates 
which should have significantly shorter decoherence time that 
the NMR or ion-trap gates. 

One more difficult 
problem of the organization of the quantum logic circuits out of 
individual qubits is the necessity to switch on and off the 
interaction between the qubits in the controlled manner. Parameters 
of the direct Coulomb interaction between the charge qubits and 
magnetic interaction between the flux qubits typically depend on 
the system geometry and can not be controlled by external voltage 
or current signals. Two possible solutions of this problem has 
been suggested so far. One is the spatial separation of the qubit 
states in the arrays of junctions \cite{ssc}, another is making 
the interaction between the qubits dependent on the ``internal'' 
qubit parameter and controlling this parameter \cite{sch2}. 
The advantage of the first approach is that it gives the possibility 
of controlling the interaction with sufficiently high accuracy, as 
needed for quantum computation. However, the difficulty of 
fabricating and controlling long arrays of junctions make simpler 
approaches like \cite{sch2} attractive for the near-term experiments.

\section{Single-electron transistor as quantum detector} 

\vspace{1ex} 

Essential part of the operation of a quantum circuit is the 
measurement process. In the case of the quantum computing circuits, 
the measurement is necessary for the read-out of the final state, 
for error-correction during operation, and, possibly, for preparation 
of the initial state. The standard detector for measurements of 
charge in systems of mesoscopic tunnel junctions is the 
single-electron tunneling (SET) transistor \cite{s1,s2,s3}. 
Over the last ten years, it has been extensively developed 
\cite{s4,s5,s6,s7,s8,s9} and used in many experiments as 
electrometer -- see, e.g., \cite{s10,s11,s12}. In this Section, we 
study the detector properties of SET transistor in a quantum 
measurement. Quantum measurements with the standard flux detector, 
a dc SQUID, are discussed in \cite{s25}. 

\begin{figure}[htb]
\setlength{\unitlength}{1.0in}
\begin{picture}(3.,1.7) 
\put(1.1,0.1){\epsfxsize=1.1in\epsfysize=1.7in\epsfbox{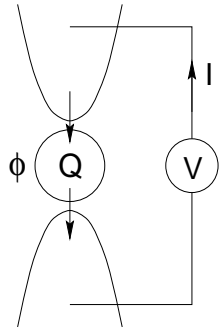}}
\end{picture}
\caption{ Diagram of the SET transistor as electrometer. 
The tunnel current $I$ (measurement output) indicated by arrows 
is controlled by electrostatic potential $\phi$ of the central 
electrode of the transistor (input signal) created by a  
measured system, and is coupled to the transistor through 
the charge $Q$ of its central electrode. }
\label{ff1} \end{figure}

SET transistor \cite{s1} is a small conductor, i.e., semiconductor 
quantum dot or small metallic island, placed between two external 
electrodes (Fig.\ \ref{ff1}). The conductor forms two tunnel junctions 
with the electrodes and acts as the central electrode of the 
transistor. Due to Coulomb charging of this electrode by tunneling 
electrons, the current $I$ through the junctions can be controlled 
by the electrostatic potential $\phi$ of the 
central electrode. Sensitivity of the current $I$ to variations of 
$\phi$ makes it possible to detect motion of charges in the vicinity 
of the transistor, and is the basis for its operation as electrometer. 
When the bias voltage $V$ across the transistor is smaller than the 
threshold $V_t$ of Coulomb blockade, the tunneling is suppressed by 
the charging energy that is required to transfer an electron in or 
out of the central electrode. In this regime, the current $I$ flows 
through the transistor only by a two-step ``macroscopic quantum 
tunneling'' (MQT) process via a virtual intermediate state below 
the Coulomb energy barrier (for review, see \cite{s16}), and is 
much smaller than the current above the Coulomb blockade 
threshold $V_t$. 

Rapid current rise at $V\simeq V_t$ makes the Coulomb blockade 
threshold the optimum operating point of the SET transistor 
as electrometer \cite{s17,s18}. Motivated by this fact, we 
analyze below characteristics of the SET electrometer as a 
quantum detector close to the Coulomb blockade threshold. 
Transport properties of the SET transistor in the vicinity of the 
threshold are determined by the crossover from MQT 
to the regular single-electron tunneling. The dominant feature of 
this crossover is quantum broadening of the transistor charge states 
by tunneling \cite{s17,s20,s21,s22,s23}. Such a broadening 
determines the shape of the current-voltage characteristics $I(V)$ 
at $V\simeq V_t$, and can be described in the simplest way in the 
regime when the temperature and ``bias energy'' $eV$ are smaller 
than the energy spacing of the single-particle states of the central 
electrode. The tunnel current $I$ flows then only through one 
single-particle state and the problem becomes equivalent to the 
standard problem of resonant tunneling through an individual 
single-particle level. Although this is not the regime of operation 
of a typical Al-based SET electrometer, it is routinely realized in 
transport through semiconductor quantum dots -- see, e.g., \cite{s24}, 
and is reached even in Al transistors \cite{s9}. 

The resonant tunneling through one single-particle level can be 
described quantitatively with the standard Hamiltonian: 
\begin{equation} 
H= \varepsilon a^{\dagger}a + \sum _{jp} ( \varepsilon_p 
c^{\dagger}_{jp}c_{jp} + T_{jp} c^{\dagger}_{jp} a + 
T_{jp}^* a^{\dagger}c_{jp} ) \, .  
\label{r1} \end{equation} 
The operators $c_{jp}$ represent electrons in the state $p$ with 
energy $\varepsilon_p$ of the $j$th electrode of the transistor, 
$j=1,2$, while operator $a$ represents electron occupying the 
resonant level. The energy $\varepsilon$ of the resonant level 
depends on the electrostatic potential $\phi$ of the central 
electrode of the transistor, $\delta \varepsilon =-e \delta \phi$.  
Potential $\phi$ controls the tunnel current $I$ through the 
transistor and serves as the input signal of the measurement. We 
assume that the measured system is coupled weakly to the transistor, 
so that the typical variations of $\phi$ and $\varepsilon$ are 
small and transistor operates as a linear detector. In the 
resonant-tunneling regime this implies that $\delta \varepsilon \ll 
eV, \, \gamma$, where $\gamma$ is the tunneling width of the 
resonant level. We also assume that the signal frequency is much 
smaller than $eV$ and $\gamma$ and treat the transistors response as 
instantaneous. 

The characteristics of a linear detector relevant for quantum 
measurement are the output noise, back-action noise of the detector 
on the measured system, and the linear response coefficient 
of transformation of the input into the output signal -- see,e.g., 
\cite{s25}. For the SET transistor, the measurement output is the 
current $I$ through the junctions. The output noise is then the 
current noise with spectral density $S_I(\omega)$, while the 
response coefficient $\lambda$ is equal to $\delta \langle I\rangle 
/\delta \phi$, where $\delta\langle I\rangle$ is 
the change of the average current $I$ due to variation $\delta \phi$ 
of the potential $\phi$. When the potential $\phi$ is produced by a 
dynamic system, e.g., a qubit, the transistor is coupled to this 
system by coupling energy $\phi Q$, where $Q=ea^{\dagger}a$ is 
the operator of the charge on the resonant level. Since the charge 
$Q$ fluctuates due to the current flow through the resonant level, 
the coupling creates a fluctuating force acting on the measured 
system that leads to dephasing of the quantum states of this system. 
Such a ``backaction dephasing'' is the fundamental aspect of a 
quantum measurement that is responsible for localization of the 
measured system in the eigenstates of the measured observable. In the 
case of SET transistor, the measured observable is the potential 
$\phi$, and intensity of the backaction dephasing is determined by 
the spectral density of the charge noise $S_Q(\omega)$. 

In the assumed regime of low signal frequency, only the zero-frequency 
values of the spectral densities of both the output and the input 
noise, as well as the spectral density $S_{IQ}(\omega)$ of the 
correlator between $I$ and $Q$ are relevant for the measurement 
problem. The zero-frequency value of $S_{IQ}(\omega)$ can be 
calculated as 
\begin{equation} 
S_{IQ}= \frac{1}{2\pi} \int dt [ \langle I(t)Q \rangle -\langle 
I \rangle \langle Q \rangle ] \, , 
\label{r2} \end{equation}
with the zero-frequency limits of $S_I(\omega)$ and $S_Q(\omega)$ 
given by similar expressions. 

Because of the simple quadratic form of the Hamiltonian (\ref{r1}) 
all these characteristics of the transistor can be found 
exactly. Partially, this was done in the theory of charge 
sensitivity of the SET electrometer in the resonant-tunneling 
regime \cite{s26,s30a}. Calculation of the charge sensitivity 
for measurements of static charges requires 
only the knowledge of the current noise $S_I$, whereas under a 
generic measurement conditions one needs to know also the backaction 
noise $S_Q$ and the correlation $S_{IQ}$. Below all spectral 
densities are calculated following Ref.\ \cite{s26}. In contrast 
to other treatments of noise properties of the resonant tunneling 
structures \cite{s27,s28} which focus on the suppression of 
shot noise, discussion of the optimum detector properties of such 
structures requires the consideration of the noise properties at 
temperatures and bias voltages smaller than the width $\gamma$ of 
the resonant level.  

The Heisenberg equations of motion for operators $c_{jp}$ and $a$ 
obtained from the Hamiltonian (\ref{r1}) are: 
\begin{equation} 
i\dot{c}_{jp}= \varepsilon_p c_{jp}-T_{jp}a\, , \;\;\; 
i\dot{a}=\varepsilon a -\sum_{jp} T_{jp}^*c_{jp} \, .  
\label{r3} \end{equation}
Integrating the first of eqs.\ (\ref{r3}) we get: 
\begin{equation}
c_{jp}(t)=\bar{c}_{jp}(t)+iT_{jp} \int^t_{-\infty} dt'
e^{-i\varepsilon_p(t-t')}  a(t') \, , 
\label{r4} \end{equation}
where $\bar{c}_{jp}$ is the ``initial value'' of $c_{jp}$ that 
describes the equilibrium occupation of electron states in the 
transistor electrodes: 
\begin{equation} 
\langle \bar{c}_{jp}^{\dagger}(t)\bar{c}_{jp}(t') \rangle= 
f_j(\varepsilon_p) e^{i\varepsilon_p(t-t') }\, . 
\label{r5} \end{equation}
Here $f_j(\varepsilon_p)$ are the Fermi distribution functions with 
the chemical potentials $\mu_j$ of the two electrodes, $\mu_1-\mu_2 
=eV>0$.    
 
The second of eqs.\ (\ref{r3}) combined with eq.\ (\ref{r4}) can be 
transformed as 
\begin{equation} 
\dot{a}=-i\varepsilon a +i\sum_{jp} T_{jp}^*\bar{c}_{jp} - \sum_{jp} 
|T_{jp}|^2 \int_0^{\infty} dt' e^{i\varepsilon_pt'} a(t-t') \, . 
\label{r6} \end{equation} 
Under the assumption that the electrodes have continuous 
spectrum of energy states, and that the densities of states $\rho_j 
(\varepsilon_p)$ together with the matrix elements of tunneling are 
constant in a wide energy range around $\varepsilon_p \simeq 
\varepsilon$, the last term in this equation reduces to 
$-\gamma a(t)$, with
\[ \gamma =\gamma_1 +\gamma_2\, , \;\;\; \gamma_j= \pi |T_j|^2 
\rho_j(\varepsilon ) \, . \]  
Here $|T_j|^2$ is the tunneling probability $|T_{jp}|^2$ averaged 
over all states with the same energy $\varepsilon_p$, and 
$\gamma_j$ is half of the tunneling rate between the $j$th electrode 
and the resonant level. After this transformation, eq.\ (\ref{r6}) 
gives: 
\begin{equation}
a(t)=\sum_{jp}\frac{ T_{jp}^* \bar{c}_{jp} (t)}{\varepsilon- 
\varepsilon_p -i\gamma } \, . 
\label{r7} \end{equation}

Equation (\ref{r7}) allows us to find the backaction noise $S_Q$: 
\begin{eqnarray}
\lefteqn{ S_Q=\frac{e^2}{2\pi} \int dt \langle a^{\dagger}(t)a 
\rangle \langle a(t)a^{\dagger} \rangle = } \nonumber \\
\frac{e^2}{\pi^2}\sum_{i,j} & & \gamma_i \gamma_j \int \frac{d\nu}{ 
[(\varepsilon- \nu)^2 + \gamma^2]^2 } f_i(\nu)(1-f_j(\nu)) \, .
\label{r9} \end{eqnarray} 
In the limit of low temperatures, $T\ll eV, \gamma$, relevant 
for transistor operation as a quantum detector, eq.\ (\ref{r9}) 
reduces to the following form:
\begin{equation}
S_Q=\frac{e^2\gamma_1 \gamma_2}{\pi^2} \int_{\mu_2}^{\mu_1} 
\frac{d\nu}{ [(\varepsilon- \nu)^2 + \gamma^2]^2 } \, . 
\label{r9a} \end{equation}
When either temperature $T$ or bias $eV$ are larger than the 
tunneling width $\gamma$, integral in eq.\ (\ref{r9}) can be taken 
treating $f$'s as constants, and gives expression for the noise $S_Q$ 
that can also be obtained from solution of a simple kinetic equation 
for the occupation probabilities of the resonant level. The same 
conclusion applies to eq.\ (\ref{r9a}) in the case of large voltages, 
$eV\gg \gamma$, when the noise $S_Q$ in the occupation of the 
resonant level is due the shot noise of the tunneling current. In 
this case, eq.\ (\ref{r9a}) gives: 
\begin{equation}
S_Q=\frac{e^2\gamma_1 \gamma_2}{2\pi\gamma^3} \, .
\label{r9b} \end{equation}

In order to find the average current $\langle I\rangle$ and the 
noise $S_I$ at low frequencies, the tunneling current $I$ through 
the transistor can be calculated in either of the junctions. 
In the first junction, 
\begin{equation} 
I= ie\sum_p (T_{1p} c^{\dagger}_{1p} a - T_{1p}^* a^{\dagger} 
c_{1p} ) \, . 
\label{r10} \end{equation} 
Using explicit expression (\ref{r7}) for $a$ in eq.\ (\ref{r4}) 
to express $c_{jp}$ in terms of $\bar{c}_{jp}$, we write the 
operator (\ref{r10}) of the current $I$ as 
\begin{equation}
I=ie (A^{\dagger} a -a^{\dagger} A) \, , 
\label{r11} \end{equation} 
where 
\begin{equation} 
A^{\dagger} \equiv \sum_p T_{1p} \bar{c}^{\dagger}_{1p} - 
i\gamma_1 \sum_{jp}\frac{ T_{jp} \bar{c}^{\dagger}_{jp} 
(t)}{\varepsilon- \varepsilon_p +i\gamma } \, . 
\label{r11a} \end{equation} 
Averaging eq.\ (\ref{r11}) over the equilibrium state of the 
electrodes as in the eq.\ (\ref{r5}), we get the standard result 
for the dc current through a resonant level, and find the 
response coefficient $\lambda \equiv \delta \langle I\rangle 
/\delta \phi$: 
\begin{equation} 
\lambda = \frac{4 e^2 \gamma_1 \gamma_2 }{\pi}\int d\nu\frac{ 
\varepsilon -\nu}{ [(\varepsilon- \nu)^2 + \gamma^2]^2 } 
(f_1(\nu)-f_2(\nu)) \, . 
\label{r12} \end{equation} 
At large bias, when $\varepsilon- \mu_2 \gg \gamma$ and $E 
\equiv \mu_1- \varepsilon \gg \gamma$, the response coefficient 
decreases, e.g., for low temperatures,      
\begin{equation} 
\lambda= \frac{2 e^2 \gamma_1 \gamma_2 }{\pi E^2} \, .  
\label{r13} \end{equation} 

The noise $S_I$ can also be found from eq.\ (\ref{r11}). Similarly 
to eqs.\ (\ref{r2}) and (\ref{r9}), we get: 
\[ S_I=\frac{e^2}{2\pi} \int dt [\langle A^{\dagger}(t)A  
\rangle \langle a(t)a^{\dagger} \rangle + \langle A(t)A^{\dagger}  
\rangle \langle a^{\dagger}(t)a \rangle - \]
\[ \langle A^{\dagger}(t)a  
\rangle \langle a(t)A^{\dagger} \rangle - \langle A(t)A^{\dagger}  
\rangle \langle a^{\dagger}(t)a \rangle ] \, . \] 
Evaluating averages in this expression with the help of eqs.\ 
(\ref{r11a}) and (\ref{r7}), we get after some algebra: 
\[  S_I=\frac{e^2\gamma_1 \gamma_2}{\pi^2} \int \frac{d\nu}{ 
[(\varepsilon- \nu)^2 + \gamma^2]^2 } \] 
\begin{equation}
[\gamma_1 \gamma_2 (f_1(\nu)(1-f_1(\nu))+f_2(\nu)(1-f_2(\nu))) + 
\label{r14} \end{equation} 
\[ ((\varepsilon- \nu)^2 + (\gamma_1- \gamma_2)^2) (f_1(\nu) 
(1-f_2(\nu))+ (1-f_1(\nu)) f_2 (\nu)) ] \, . \]
Equation (\ref{r14}) gives the current noise at arbitrary 
temperatures and bias voltages. In the limit of small 
temperatures, it can be written as 
\begin{equation}
S_I=\int_{\mu_2}^{\mu_1} d\nu S_I(\nu) \, , 
\label{r15} \end{equation}  
\[ S_I(\nu)= \frac{e^2\gamma_1 \gamma_2}{\pi^2} \frac{(\varepsilon- 
\nu)^2 + (\gamma_1- \gamma_2)^2}{ [(\varepsilon- \nu)^2 + 
\gamma^2]^2 } \, . \] 
Note that the partial contribution $S_I(\nu )d\nu$ to the current 
noise (\ref{r15}) from an energy interval $d\nu$ around some energy 
$\nu$ is given by the standard general equation for the noise of a 
mesoscopic conductor -- see, e.g., \cite{s30}, $S_I(\nu) d\nu =
(e^2 d\nu/4\pi^2)D(1-D)$, where the transparency $D$ of the 
resonant-tunneling structure is energy-dependent: $D=4\gamma_1 
\gamma_2/((\varepsilon- \nu)^2 + \gamma^2)$, and the width $d\nu$ 
of the energy interval acts as the bias energy $eV$.  
   
In the large-bias limit, eq.\ (\ref{r15}) describes current noise  
associated with the incoherent sequential tunneling through the 
resonant level \cite{s27,s28,s26}:  
\begin{equation}
S_I=\frac{e\langle I\rangle}{2\pi}( 1-\frac{2\gamma_1\gamma_2 
}{\gamma^2}) \, .
\label{r16} \end{equation}
where the average current through the level is $\langle I\rangle 
=2e\gamma_1 \gamma_2/\gamma$. In structures that are not highly 
asymmetric, $\gamma_1\sim \gamma_2$, eq.\ (\ref{r16}) 
describes suppression of the current noise in comparison with the 
regular shot noise $e\langle I\rangle/2\pi$. The previous 
discussion shows that this suppression has two correct 
interpretations that appear to be quite different. One explanation 
relates the suppression to the quantum Fermi correlations in the 
tunneling flow of electrons through the structure described by 
the transparency-dependent suppression factor $(1-D)$. In another 
interpretation, the suppression is brought about by the classical 
correlations in the process of electron transfer through the 
structure, which break this process into two distinct steps, 
tunneling onto the resonant level and out of it.  

Finally, following the same steps that lead to eq.\ (\ref{r14}) 
for the current noise, we calculate the current-charge correlator 
$S_{IQ}$. From eqs.\ (\ref{r2}) and (\ref{r11}), and the definition 
of $Q$ we have: 
\[ S_{IQ} =\frac{ie^2}{2\pi} \int dt [\langle A^{\dagger}(t)a 
\rangle \langle a(t)a^{\dagger} \rangle - \langle A(t)a^{\dagger} 
\rangle \langle a^{\dagger}(t)a \rangle ]\, . \]
Making use of eqs.\ (\ref{r7}) and (\ref{r11a}) to calculate the 
averages in this expression we obtain: 
\[ S_{IQ} =\frac{e^2\gamma_1 \gamma_2}{\pi^2} \int \frac{d\nu}{ 
[(\varepsilon- \nu)^2 + \gamma^2]^2 } [\gamma_1 f_1(\nu)(1- 
f_1(\nu))-\] 
\[ \gamma_2 f_2(\nu)(1-f_2(\nu)) + (\gamma_2- \gamma_1 - 
i(\varepsilon- \nu) ) f_1(\nu) (1-f_2(\nu))+\ \] 
\begin{equation}
(\gamma_2- \gamma_1 +i(\varepsilon- \nu) )
(1-f_1(\nu)) f_2 (\nu)) ] \, . 
\label{r17} \end{equation} 
At $T\ll \gamma,eV$, this equation reduces to 
\begin{equation}   
S_{IQ} =\frac{e^2\gamma_1 \gamma_2}{\pi^2} \int_{\mu_2}^{\mu_1} 
d\nu \frac{\gamma_2 - \gamma_1 - i(\varepsilon- \nu)}{ 
[(\varepsilon- \nu)^2 +  \gamma^2]^2 }\, .
\label{r18} \end{equation}

Comparing eqs.\ (\ref{r9a}), (\ref{r15}), and (\ref{r18}) for the 
low-temperature noise spectral densities, and the low-temperature 
limit of eq.\ (\ref{r12}) for the response coefficient, we can see 
that the partial contributions to these functions from tunneling 
of electrons with a certain energy $\nu$ satisfy the generic 
relations of a linear quantum-limited detector \cite{s25}: 
\begin{equation}
S_I(\nu) S_Q(\nu)= |S_{IQ}(\nu)|^2 \, , \;\;\;\; \mbox{Im}S_
{IQ}(\nu)=-\lambda(\nu)/4\pi \, ,
\label{r19} \end{equation}
while, similarly to other types of detectors, $\mbox{Re}S_{IQ}$ 
characterizes the detector asymmetry. 
At small bias voltages, $eV\ll \gamma$, when the tunnel current is 
determined by tunneling in the narrow energy range, the actual 
values of $\lambda$ and all spectral densities coincide with 
$\lambda(\nu)$, $S_I(\nu)$, etc. According to generic arguments 
presented in Ref.\ \cite{s25}, eqs.\ (\ref{r19}) mean that the 
SET transistor in this regime is a quantum-limited detector reaching 
the fundamental limit of the energy sensitivity $\epsilon$:  
\begin{equation} 
\epsilon \equiv \frac{2\pi }{\lambda(\nu) } [S_I(\nu) S_Q (\nu) - 
(\mbox{Re}S_{IQ}(\nu))^2]^{1/2} =\hbar/2 \, . 
\label{r20} \end{equation}

Such a quantum detector also has an ultimate signal-to-noise ratio 
$R$ for a continuous weak measurement of a qubit. If the qubit 
oscillates between its two basis states, the signal-to-noise ratio 
can be defined as the ratio of the amplitude of the oscillation line 
in the output spectrum of the detector to the output noise. The 
ratio $R$ can be shown \cite{s25} to be closely related to the 
energy sensitivity (\ref{r20}): 
\[ R= \frac{\lambda(\nu)^2}{4\pi^2 S_I(\nu) S_Q (\nu)} \, .\]
For a symmetric detector with $\mbox{Re}S_{IQ}(\nu)=0$ (i.e., in the 
case of SET transistor, when $\gamma_1=\gamma_2$), the signal-to-noise 
ratio $R$ can be written as $(\hbar/\epsilon)^2$ and is limited by 
4.\cite{s31} This fundamental signal-to-noise ratio means that the 
detector does not add more dephasing into the qubit dynamics than 
the minimum required by the acquisition of information about the 
state of the system.\cite{s32,s25} 

The small bias condition necessary for the SET transistor to operate 
as a quantum-limited detector with the energy sensitivity $\hbar/2$ 
limits the absolute value of the transistor output (the current $I$). 
In order to see to what extent the transistor properties are degraded 
by increasing bias, we consider the case when the ``collector'' 
electrode absorbs all tunneling electrons, $\varepsilon 
-\mu_2\gg \gamma$. Detector characteristics of the transistor, most 
importantly energy sensitivity $\epsilon$, depend then on the 
position of the resonant energy level relative to the chemical 
potential of the ``emitter'' electrode characterized by $E=\mu_1- 
\varepsilon$. Integrating eqs.\ (\ref{r9a}), (\ref{r15}), 
(\ref{r18}), and (\ref{r12}) we find the noise spectral densities 
as functions of $E$: 
\[ S_I=\frac{e^2 \gamma_1 \gamma_2 }{\pi^2\gamma^3}[(\gamma_1^2 + 
\gamma_2^2)(\frac{\pi}{2} +\tan^{-1}z) -2 \gamma_1 \gamma_2 
\frac{z}{1+z^2} ] \, ,\]
\begin{equation}
S_Q=\frac{e^2\gamma_1 \gamma_2}{\pi^2\gamma^3} [\frac{\pi}{2} 
+\tan^{-1}z +\frac{z}{1+z^2} ]  \, , \;\;\; z\equiv E/\gamma \, ,
\label{r21} \end{equation}  
\[ \mbox{Re} {S}_{IQ}= (\gamma_1-\gamma_2) S_Q, \;\;\; 
-\mbox{Im} {S}_{IQ}= \frac{\lambda}{4\pi}= \frac{e^2 \gamma_1 
\gamma_2 }{2\pi^2\gamma^2} \frac{1}{1+z^2} \, .\] 

\begin{figure}[htb]
\setlength{\unitlength}{1.0in}
\begin{picture}(3.,2.3) 
\put(0.2,0.0){\epsfxsize=2.7in\epsfysize=2.3in\epsfbox{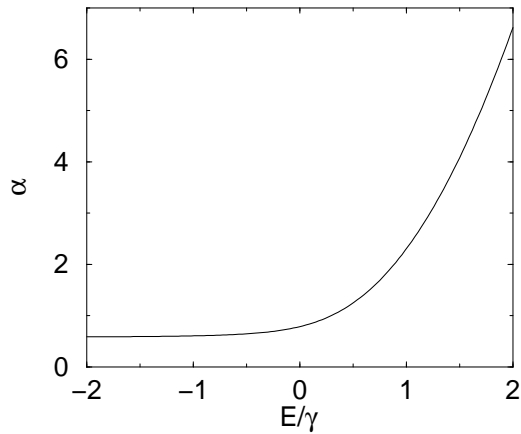}}
\end{picture}
\caption{Low-temperature energy sensitivity \protect (\ref{r22}) of 
the SET transistor in the resonant-tunneling regime in units of 
$\hbar$, $\alpha \equiv \epsilon/\hbar$, as a function of the bias 
energy $E$ characterizing position of the resonant level relative 
to the chemical potential of the emitter electrode. }
\label{ff2} \end{figure}

Using eqs.\ (\ref{r21}) we obtain that the energy sensitivity 
$\epsilon = 2\pi [S_IS_Q -(\mbox{Re}S_{IQ} )^2]^{1/2}/\lambda$ 
of SET transistor in the resonant-tunneling regime is 
independent of the degree of the transistor asymmetry 
$\gamma_1/\gamma_2$: 
\begin{equation}  
\epsilon= \frac{\hbar}{2} [(1+z^2)^2(\frac{\pi}{2} +\tan^{-1}z)^2 
-z^2]^{1/2} \, .  
\label{r22} \end{equation}
Equation (\ref{r22}) is plotted in Fig.\ \ref{ff2}. We see that 
the behavior of $\epsilon$ as a function of $E$ changes from 
being nearly-constant for $E<0$ to rapid increase with $E$ at 
$E>0$. This rapid increase reflects suppression of the response 
coefficient $\lambda$ (\ref{r13}) with increasing bias. 
Quantitatively, eq. (\ref{r22}) gives for $|E|\gg \gamma$: 
\begin{equation} 
\epsilon=\hbar \left\{ \begin{array}{l} \pi E^2/2\gamma^2, 
\;\; E>0, \\  1/\sqrt{3},\;\; E<0. \end{array} \right.  
\label{r23} \end{equation}
This equation shows that in the regime of quantum tunneling, $E<0$, 
the energy sensitivity comes very close to the fundamental limit 
$\hbar/2$. It does not quite reach it because of the energy 
dependence of the tunneling rate.  

\begin{figure}[htb]
\setlength{\unitlength}{1.0in}
\begin{picture}(3.,2.5) 
\put(0.2,0.0){\epsfxsize=2.9in\epsfysize=2.4in\epsfbox{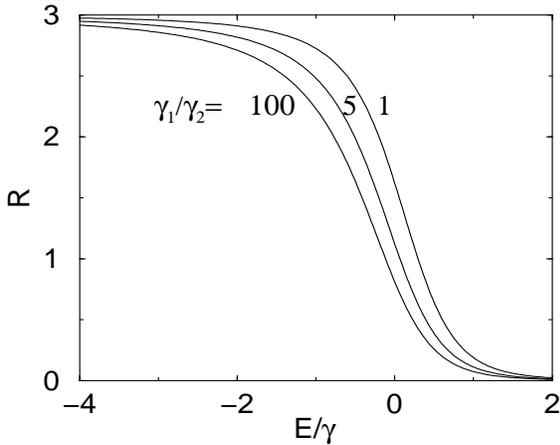}}
\end{picture}
\caption{Signal-to-noise ratio $R$ for continuous weak measurement 
of a qubit with the SET transistor in the resonant-tunneling regime.  
$R$ is shown as a function of the bias energy $E$ for various 
degrees of the transistor asymmetry. }
\label{ff3} \end{figure}

Figure \ref{ff3} shows the signal-to-noise ratio $R=\lambda^2/ 
(4\pi^2S_IS_Q)$ for continuous measurement of a qubit using the SET 
transistor in the resonant-tunneling regime that is calculated from 
eqs.\ (\ref{r21}). In contrast to $\epsilon$, $R$ 
depends on the transistor asymmetry, but the magnitude of variation 
of $R$ with $\gamma_1/\gamma_2$ is small, and regardless of 
$\gamma_1/\gamma_2$, the signal-to-noise ratio approaches 3 at 
large negative $E$. In the symmetric case, $\gamma_1/\gamma_2=1$, 
the signal-to-noise ratio is given by the general relation 
$R=(\hbar/\epsilon)^2$ with energy sensitivity (\ref{r22}).  

To conclude, we see that the SET transistor in the 
resonant-tunneling regime biased at the Coulomb blockade threshold 
should operate at low temperatures as a quantum-limited detector. 
It has energy sensitivity $\hbar/\sqrt{3}$ and signal-to-noise 
ratio 3 for continuous weak measurement of a qubit. In this regime, 
transistor introduces the noise in the dynamics of the measured 
system that is only marginally larger than the fundamental noise 
required by the quantum mechanics of measurement. This means that 
for the optimally-biased transistor the time of the backaction 
dephasing is equal to the measurement time. This conclusion 
differs from the results of the previous analysis of the SET 
transistor at large bias voltages \cite{s19}, where the 
transistor generates strong shot noise leading to the backaction 
dephasing that is much stronger than the fundamental minimum. 

The author would like to acknowledge useful discussions with A.N. 
Korotkov. This work was supported by ARO and AFOSR.


\begin{references}
\bibitem{s30a} The charge sensitivity discussed in Ref.\ \cite{s26} 
was expressed through the energy sensitivity $\delta\varepsilon$ 
of the tunnel current to variations of the position of the resonant 
level. Note that this energy sensitivity is different from the more 
standard energy sensitivity $\epsilon$ (\ref{r20}) studied in this 
work.
\end{references}
\end{document}